\documentclass[%
11pt,
twoside,
superscriptaddress,
nofootinbib,
amsmath,
amssymb,
aps,
prd,
]{revtex4-2}

\makeatletter
\def\tagform@#1{(\textcolor{red}{#1})}
\makeatother
\usepackage[colorlinks=true,linkcolor=red,citecolor=red,urlcolor=red]{hyperref}

\pdfoutput=1

\usepackage{tikz}
\usetikzlibrary{calc,trees,positioning,arrows,chains,shapes.geometric,decorations.pathreplacing,decorations.pathmorphing,shapes,matrix,shapes.symbols}
\usepackage{bm}
\usepackage{MnSymbol}
\usepackage{amsmath}
\usepackage{graphicx}
\usepackage{standalone}
\usepackage{subcaption}
\usepackage{feynmp}
\usepackage{hyperref}
\usepackage{bbold}
\usepackage{cancel}
\usepackage{eufrak}
\usepackage{upgreek}
\usepackage{tikz}
\usepackage{cancel}
\usepackage{empheq}
\usepackage[compat=1.1.0]{tikz-feynman}
\usepackage{stmaryrd,scalerel}
\usepackage{enumitem}
\usepackage{float}
\usepackage{color} 
\usepackage{etex}
\usepackage{morefloats}
\usepackage{lipsum}

\def\beq{\begin{equation}}
\def\eeq{\end{equation}}

\def\bea{\arraycolsep .1em \begin{eqnarray}}
\def\eea{\end{eqnarray}}

\def\!!!{\stackrel{!}{=}}

\def\S{ \mathcal{S}}

\def\eq#1{(\ref{#1})}

\def\s0#1#2{\mbox{\small{$ \frac{#1}{#2} $}}}
\def\0#1#2{\frac{#1}{#2}}

\def\grgl{\:\hbox to -0.2pt{\lower2.5pt\hbox{$\sim$}\hss}{\raise3pt\hbox{$>$}}\:}
\def\klgl{\:\hbox to -0.2pt{\lower2.5pt\hbox{$\sim$}\hss}{\raise3pt\hbox{$<$}}\:}

\begin{document}


\title{\color{blue} \bf Classical Renormalization Group Equations for General Relativity}

\author{F. Gutiérrez}
\email{fgutierrez@fing.edu.uy}
\affiliation{IFFI, Universidad de la Rep\'ublica, J.H.y Reissig 565, 11300 Montevideo, Uruguay}

\author{K. Falls}
\email{kfalls@fing.edu.uy}
\affiliation{IFFI, Universidad de la Rep\'ublica, J.H.y Reissig 565, 11300 Montevideo, Uruguay}

\author{A. Codello}
\email{alessandro.codello@unive.it}
\affiliation{DMSN, Ca'\ Foscari University of Venice, Via Torino 155, 30172 - Venice, Italy}
\affiliation{IFFI, Universidad de la Rep\'ublica, J.H.y Reissig 565, 11300 Montevideo, Uruguay}

\date{\today}

\begin{abstract}\vspace{0.25cm}
In a companion paper {\tt arXiv:2510.27676}, we introduced a non-perturbative classical renormalisation group (RG) flow equation as a novel method for treating strongly interacting problems in general relativity, with a prominent application to the two-body problem. While we demonstrated that it reproduces perturbation theory, via the Post-Minkowskian (PM) expansion, and its computational efficiency in reproducing the 1PN Post-Newtonian action, its derivation was heuristic. In this work, we place this flow equation on a firm formal foundation. In particular, we demonstrate that a Legendre transform maps the classical analogue of the Polchinski equation precisely to our classical RG equation. This establishes a duality between equivalent, exact RG equations for the gravitational effective action. The result, combined with the successful applications in {\tt arXiv:2510.27676}, solidifies the classical RG framework as a powerful and rigorous new approach to the general relativistic two-body problem and gravitational wave physics.
\end{abstract}

\maketitle
\section{Introduction}
\vspace{-0.2cm}

The two-body problem in general relativity, a cornerstone for predicting gravitational wave observables, is fundamentally a strong-field, non-perturbative phenomenon. While the Post-Minkowskian (PM) and Post-Newtonian (PN) expansion schemes have been highly successful \cite{Einstein:1938yz,Buonanno:1998gg,Blanchet:2002gz,Jaranowski:1997ky,Poisson:2014book}, a framework that seamlessly interpolates between weak and strong coupling remains a central theoretical goal.
%
In a companion paper \cite{Gutierrez:2025rqp} we have introduced a non-perturbative classical RG flow equation as a novel way to treat strongly interacting problems in general relativity -- prominently the two-body problem. We have first shown that this equation is perturbatively exact -- in the sense that it reproduced the PM expansion (we gave explicitly the first three orders but the procedure is easily generalisable); second we have shown how the flow equation can be used to reproduce the 1PN action with a minimal amount of computations -- and in particular without the need to employ the gravitational three vertex at all.
\\

Our derivation of the flow equation in \cite{Gutierrez:2025rqp} -- which we briefly review here -- was heuristic: we took the standard route of making the RG improvement of the leading order perturbative correction -- in the context of statistical and quantum field theory this procedure leads to the well known Morris-Wetterich flow equation \cite{Wetterich:1992yh,Morris:1993qb} (see \cite{Dupuis:2020fhh} for a review). 
In the present setting, we focus on the leading PM contribution to the effective action
\begin{equation}\label{1PM}
S_{\mathrm{eff}} = S_{\mathrm{pp}} 
- \frac{\kappa}{2} S^{a}_{\mathrm{pp}} (\Delta^{-1})_{ab} S^{b}_{\mathrm{pp}} 
+ \mathcal{O}(\kappa^2) \, ,
\end{equation}
where we are using the DeWitt index notation summarised below.
Above $\Delta = S^{(2)}_g[\bar{g}]$ denotes the inverse propagator evaluated on a solution of the vacuum Einstein equations for which $S_g^{(1)}[\bar{g}]=0$, such as Minkowski spacetime $\bar{g}_{\mu\nu}=\eta_{\mu\nu}$ and $S_{\mathrm{pp}}$ is the particle action\footnote{Here and throughout $F^{(n)}$ denotes the $n$th functional derivative of a functional $F$ with respect to the metric $g_{\mu\nu}$.}.
%
Then we introduced a (Lorentzian) infrared regulator $R_k$ in the propagator, replacing $S_{\rm g}^{(2)} \to S_{\rm g}^{(2)} + R_k$, while keeping the spacetime metric fully arbitrary, $\bar{g} \to g$. 
Moreover, we promoted the effective action to be both scale and field dependent, $S_{\mathrm{eff}} \to S_k[g]$. 
Under these assumptions \eqref{1PM} becomes:
\begin{equation}\label{1PMb}
 S_k[g] =  S_{\mathrm{pp}}[g] -\frac{\kappa}{2}S^{a}_{\mathrm{pp}}[g] (S^{(2)}_{\rm g}[g]+R_{k})^{-1}_{ab}S^{b}_{\mathrm{pp}}[g] + \mathcal{O}(\kappa^2) \,.
\end{equation}
Setting $R_k \to 0$ and $g_{\mu\nu} = \bar{g}_{\mu\nu}$ we return to \eqref{1PM}. The cutoff $R_k$ is chosen to vanish for $k=0$ and to diverge for $k \to \infty$. Thus, we have that 
\begin{equation}
\lim_{k \to \infty} S_k = S_{\mathrm{pp}}\,,
\end{equation}
and at $k=0$ the action $S_k$ gives the effective action $S_{\rm eff}$. Furthermore, $S_{k=0}[\bar{g}] =S_{\rm eff}$ to first order in $\kappa$.
Now the RG improvement: one takes a derivative of \eqref{1PMb} with respect to $k$ and promotes $S^{a}_{\mathrm{pp}}$ to be the full $S^{a}_k$ at scale $k$. So we get the final result for the classical flow equation\footnote{Where the dot ``$\cdot$'' indicates the contraction of indices and a spacetime integral.}:
\begin{equation} \label{eq:flowWett}
\boxed{\; \partial_k S_k[g] =  - \frac{\kappa}{2} S_{k}^{(1)}[g] \cdot  \partial_k G_k[g] \cdot S_{k}^{(1)}[g]\;}
\end{equation}
 where $G_k[g] = (S_{\rm g}^{(2)}[g] + R_k)^{-1}$ is the IR regulated gravitational propagator evaluated for an arbitrary metric $g_{\mu\nu}$. The action for the point-particle enters as the initial condition while $S_{k=0}[\bar{g}] =S_{\rm eff}$ to {\it all orders} in $\kappa$. 
\\

Clearly, the leading order RG improvement does not guarantee that the flow equation obtained is exact, and there are many examples in which this is not the case \cite{Litim:2001hk,Litim:2002hj,Litim:2002xm}.
To demonstrate that \eqref{eq:flowWett} is indeed exact, we will show that it is related to the classical Polchinski equation \cite{Polchinski:1983gv} by a Legendre transform \cite{Morris:1993qb, Morris:2015oca}.  This establishes that, in a certain sense, \eqref{eq:flowWett} can be viewed as the ``Classical Morris-Wetterich equation'', for the average effective action $S_k$. This result -- together with the application mentioned in the previous paragraph --  establishes our formalism as a firm new theoretical approach to the two-body problem and gravitational wave physics.
\\

This paper is organised as follows: In Sec.~\ref{Sec:2}, we derive the classical Polchinski equation for the gravitational two-body problem. In Sec.~\ref{Sec:3}, we perform the Legendre transform and establish the duality that relates it to our flow equation \eqref{eq:flowWett}. We conclude in Sec.~\ref{Sec:4} by discussing the implications of this formal foundation and future applications. In the appendix, we re-derive the PM expansion from the classical Polchinski formalism of Sec.~\ref{Sec:2}.
\\

\section{Classical Polchinski equation}
\label{Sec:2}
\vspace{-0.2cm}

Our starting point is the action for general relativity coupled to two point-particles 
\beq \nonumber
S_{\rm tot}[g,x_1,x_2] = \frac{1}{\kappa} S_{\mathrm{g}}[g] + S_{\mathrm{pp}}[g,x_1,x_2] \,,
\eeq
where $\kappa = 32 \pi G_N$ is proportional to Newton's constant $G_N$ and
 \beq
 \frac{1}{\kappa}S_{\rm g}[g] = \frac{2c^4}{\kappa} \int dt \, d^3\mathbf{x} \,\sqrt{-g}\,R  + S_{\rm gf}
 \eeq
is the Einstein--Hilbert action with a gauge fixing term added.
We denote by $\bar{g}$ a solution of the \textit{background equation of motion}
\beq \label{eq:BEoM}
\frac{\delta S_{\mathrm{g}}}{\delta g_{\mu\nu}(x)}[\bar{g}] = 0 \,,
\eeq
and define\footnote{To simplify expressions, we introduce DeWitt indices such that lower case Latin indices include the spacetime indices and the coordinates e.g., $g_a = g_{\mu\nu}(x)$.
Then the $n$th functional derivative of a functional $F$ is denoted by 
$
F^{a \dots a_n} := \frac{\delta^n F}{\delta g_{a_1} \dots \delta g_{a_n}} \nonumber
$
Repeated Latin indices imply a sum over spacetime indices and an integral over spacetime, e.g.
$
g_a J^a = \int d^4 x g_{\mu\nu}(x) J^{\mu\nu}(x)\,. 
$.}
\beq \nonumber
\Delta^{ab} = \frac{\delta^2 S_{\rm g}}{\delta g_a \delta g_b}[\bar{g}]
\eeq
as the corresponding inverse propagator. 
For example $\bar{g}$ could be the Minkowski metric $\eta$.
Expanding around $\bar{g}$, we have  
\beq \nonumber
S_{\rm g}[\bar{g} + h] = 
 \frac{1}{2} h_a \Delta^{ab} h_b 
+ \mathcal{O}(h^3) \,,
\eeq
where $h_a = h_{\mu\nu}(x) = g_{\mu\nu}(x) - \bar{g}_{\mu\nu}(x) $ denotes a fluctuation field.  
If the configuration $g^a = g^a_*[x_1,x_2]$ satisfies  
\beq \nonumber
\frac{\delta S_{\rm tot}}{\delta g_a(x)}[g_*[x_1,x_2],x_1,x_2] = 0 \,,
\eeq
we can define the \textit{effective action} as  
\beq \label{eq:Seff_0}
S_{\rm eff}[x_1,x_2] = S_{\rm tot}[\tilde{g}[x_1,x_2],x_1,x_2] \,.
\eeq
In general, this effective action is difficult to compute directly, and progress 
is usually made by performing an expansion in $\kappa$.

Our goal is to construct a scale dependent action $\S_k[h,x_1,x_2]$ such that  
\begin{equation} \label{eq:Seff_k_0}
S_{\rm eff}[x_1,x_2] = \lim_{k\to 0} \S_k[0,x_1,x_2] \,,
\end{equation}
with Wilsonian effective action $\S_k[h,x_1,x_2]$ determined through a flow equation. Specifically, the flow is 
governed by the classical Polchinski equation \cite{Polchinski:1983gv}:  
\beq \label{eq:floweq}
\boxed{\; 
 \partial_k \S_k[h] = - \frac{\kappa}{2}\, 
\S_{k}^{(1)}[h] \cdot 
  \partial_k\mathcal{G}_k  \cdot  
\S_{k}^{(1)}[h] 
\;}
\eeq
where the usual one-loop term in the quantum Polchinski equation is neglected since we work in the classical limit $\hbar \to 0$.
In \eq{eq:floweq} the IR regularised propagator is given by  \beq \label{eq:cebra}
\mathcal{G}_k= G_k[\bar{g}]=\frac{1}{\Delta + R_k}\,.
\eeq
The flow equation has the initial condition  
\beq \label{eq:IC}
\lim_{k\to \infty} \S_k[h,x_1,x_2] 
= S[\bar{g}+h,x_1,x_2] 
- \frac{1}{2\kappa} h^a \Delta_{ab} h^b \,,
\eeq
which guarantees \eq{eq:Seff_k_0}. As is standard, we have subtracted the free action in the above equation to define $\S_k$ as the interacting part of the Wilsonian effective action.
To clarify this construction, let us introduce a UV-regularized propagator  
\beq \nonumber
P_k = \frac{1}{\Delta} - \frac{1}{\Delta + R_k} \,.
\eeq
and note that  
\beq \nonumber
\partial_k \mathcal{G}_k = -  \partial_k P_k \,,
\eeq
Then, in terms of the \textit{total action}
\beq \label{tot action}
\S^{\rm tot}_k[h,x_1,x_2] = \frac{1}{2 \kappa } h \cdot P^{-1}_k \cdot h + \S_k[h, x_1,x_2] \,,
\eeq
which includes the free action,
the flow equation \eqref{eq:floweq} takes the form  
\beq \label{FlOw}
 \partial_k \S^{\rm tot}_k = \kappa \, \S^{{\rm tot}(1)}_k \cdot 
\frac{ \partial_k P_k}{2}\cdot 
\big(\S^{{\rm tot}(1)}_k - \frac{2}{\kappa}P^{-1}_k \cdot h \big) \,.
\eeq
Note that $P_k$ gives us a measure of how integrated out the metric fluctuations are. In particular, when $k\to \infty$ we have $P_k \to \Delta^{-1}$, indicating that we have not integrated out any metric fluctuations. When $k= 0$ we instead have $P_0 = 0$ and hence the metric fluctuations do no longer propagate and thus have been integrated out. On the other hand the form of \eq{FlOw} guarantees that we are only changing the description of the physics. To see this clearly let's then define the $k$ dependent solution to the equation of motion $\tilde{h}_k[x_1,x_2]$ by
\beq \nonumber
\frac{\delta  \S^{\rm tot}_k}{\delta h^a(x)}[\tilde{h}_k[x_1,x_2],x_1,x_2] =0\,.
\eeq
It then follows from this definition that 
\begin{align*}
\frac{d}{dk} \S^{\rm tot}_k[\tilde{h}_k[x_1,x_2],x_1,x_2] &= \partial_k \S^{\rm tot}_k[\tilde{h}_k[x_1,x_2],x_1,x_2] 
\\
&+ \partial_k \tilde{h}_k[x_1,x_2] \cdot \S^{{\rm tot}(1)}_k[\tilde{h}_k[x_1,x_2],x_1,x_2]  = 0\,.
\end{align*}
where the first term is zero since it is given by \eq{FlOw} evaluated as $h = \tilde{h}_k$ and hence proportional to $\S^{{\rm tot}(1)}_k[\tilde{h}_k]= 0$, and the second term vanishes similarly via the flow equation \eqref{FlOw}. 
Thus, the on-shell effective action is an RG invariant.
Its value can be determined from the initial condition $ \S^{\rm tot}_k[h]|_{k \to \infty} =S[\bar{g}+h]$ which implies that
\beq \nonumber
S_{\rm eff}[x_1,x_2] =   \S^{\rm tot}_k[\tilde{h}_k[x_1,x_2],x_1,x_2] 
\eeq
which follows from \eqref{eq:Seff_0} which defines $S_{\rm eff}[x_1,x_2]$, .
This means the effective action can be computed from $\S^{\rm tot}_k$ at any scale $k$. However, for $k \neq 0$ this is difficult since we have to solve for $\tilde{h}_k[x_1,x_2]$. 
On the other hand, by using the condition $\lim_{k\to 0} P_k \to 0$ in \eqref{tot action} we have that the solution always vanishes
\beq \nonumber
\lim_{k\to 0} \tilde{h}_k[x_1,x_2]= -  2 \kappa\,  \lim_{k\to 0}  P_k \cdot \S_k^{(1)}[\tilde{h}_k, x_1,x_2] =  0\,.
\eeq
It follows that when $k=0$ we can set the fluctuations to zero and hence we have \eqref{eq:Seff_k_0}. The Wilsonian interpretation should be clear: at $k=0$ we have integrated out all fluctuations of the metric as characterised by the vanishing of the propagator $P_k$. So when $k =0$ the equation of motion for $h_{\mu\nu}$ is trivial, and we again easily obtain the effective action.

Note that, although \eqref{eq:floweq} is somewhat similar to \eqref{eq:flowWett}, it differs in two important ways. Firstly, the propagator in \eqref{eq:floweq} is evaluated on the background rather than the full propagator $G_k[g]$. Secondly, the initial condition for $\S_k$ \eqref{eq:IC} involves the action $S_{\rm g}$, e.g., the pure gravity action.
\\

\section{Classical average action}
\label{Sec:3}
\vspace{-0.2cm}

Although the classical Polchinski equation for the Wilsonian effective action $\S_k$ is exact, it requires $\S_k$ to contain a pure gravity part. Consequently, $\S^{(1)}_k$ does not correspond to a running energy--momentum tensor. To make this explicit, we decompose $\S_k$ as
\beq \label{eq:shat}
\S_k[h,x_1,x_2] = \frac{1}{\kappa} \S_{-1,k}[h] + \S_{\mathrm{pp}, \, k}[h,x_1,x_2] \, ,
\eeq
where $\S_{-1}^{k}$ denotes the pure gravity sector, which is of order $\kappa^{-1}$. The term $\S_{\mathrm{pp}, \, k}[h,x_1,x_2]$ contains the terms involving the matter fields $x_1,x_2$ (point-particles).
It then follows from \eqref{eq:floweq} and \eqref{eq:shat} that
\begin{align*}
\tfrac{1}{\kappa}\partial_k\S_{-1}+\partial_k \S_{\mathrm{pp}}
&= \S_{\mathrm{pp}}^{(1)} \cdot \frac{\partial_kR_k}{(\Delta + R_k)^2}  \cdot \S_{-1}^{(1)} 
+ \tfrac{1}{2\kappa}\S_{-1}^{(1)} \cdot \frac{\partial_kR_k}{(\Delta + R_k)^2}\cdot \S_{-1}^{(1)}
\\
&+ \frac{\kappa}{2}\S_{\mathrm{pp}}^{(1)} \cdot \frac{\partial_kR_k}{(\Delta + R_k)^2} \cdot \S_{\mathrm{pp}}^{(1)} \, . \nonumber
\end{align*}
Considering the terms of order $\kappa^{-1}$, we observe that the pure gravity part $\S_{-1}[h]$ satisfies the flow equation
\beq \label{eq:rana}
 \partial_k \S_{-1} = \frac{1}{2}\, 
\S_{-1}^{(1)} \cdot 
\frac{\partial_kR_k}{(\Delta + R_k)^2} \cdot 
\S_{-1}^{(1)} \,.
\eeq
and the matter-dependent part $\S_{\mathrm{pp}}[h,x_1,x_2]$ satisfies 
\beq \label{eq:sflow}
 \partial_k \S_{\mathrm{pp}} = \S_{\mathrm{pp}}^{(1)} \cdot \frac{\partial_kR_k}{(\Delta + R_k)^2} \cdot \S_{-1}^{(1)}
+ \frac{\kappa}{2} \S_{\mathrm{pp}}^{(1)} \cdot \frac{\partial_kR_k}{(\Delta + R_k)^2} \cdot \S_{\mathrm{pp}}^{(1)} \, .
\eeq
Following the spirit of \cite{Morris:1993qb,Morris:2015oca} where the Wetterich-Morris equation was obtained from the Polchinski equation,  we now perform a Legendre transformation of the {\it pure gravity part} of the action $\S_k$ to obtain a $k$--independent functional $S_{-1}[H]$. In particular, we define
\beq \label{eq:Gammahat}
\boxed{\;S_{-1}[H] = \frac{1}{2} H_a \Delta^{ab} H_b + \S_{-1}[h] 
- \frac{1}{2}(h- H)_a (\Delta+ R)^{ab} (h -H)_b \;}
\eeq
Since this is a Legendre transformation, the following identities hold:
\beq \label{eq:ballena}
\S_{-1}^{a}[h] =  (\Delta+ R)^{ab} (h-H)_b \, ,
\eeq
\beq \label{eq:raya}
S_{-1}^{a}[H] +R^{ab}H_b =   (\Delta+ R)^{ab} h_b \, ,
\eeq
which allow us to express $H$ in terms of $h$, and vice versa.  
Taking one functional derivative of \eqref{eq:raya}, we obtain
\[
S_{-1}^{ab}[H] + R^{ab} =  (\Delta+ R)^{a c} \frac{\delta h_c}{\delta H_b} \, ,
\]
which can be rewritten as
\beq \label{eq:hH}
\frac{\delta h_c}{\delta H_b} =  \mathcal{G}_{cd} \, (S_{-1}^{(2)}[H] +R)^{db} \, , 
\qquad
\frac{\delta H_c}{\delta h_b} = \left( \big(S_{-1}^{(2)}[H] +R\big)^{-1}  \right)_{cd}  \left(\mathcal{G}^{-1}\right)^{db} \, ,
\eeq
where we have used the fact that the two expressions are inverses of each other.
%
Differentiating \eqref{eq:Gammahat} with respect to $k$ at constant $H$, we find
\[
 \partial_k  S_{-1}[H] 
=   \partial_k \S_{-1}[h] -  \tfrac{1}{2}(h- H)^a ( \partial_k  R)_{ab} (h -H)^b \, ,
\]
where we have used that $h=h_k[H]$ maximizes the functional $\S_{-1}[h] 
- \frac{1}{2}(h- H)_a (\Delta+ R)^{ab} (h -H)_b$. However, using \eqref{eq:ballena} and \eqref{eq:rana}, the two terms cancel, yielding
\begin{equation}
k\partial_k  S_{-1}[H] = 0 \, .
\end{equation}
If we rearrange \eq{eq:raya} for $h_a = h_a[H]$ and use both \eq{eq:ballena} and \eq{eq:raya},
it follows that the $k$-derivative of the former is given by
\beq \label{eq:hflow}
 \partial_k h_b[H] = -  \mathcal{G}_{ba} \dot{R}^{ac} \mathcal{G}_{cd} \S_{-1}^{d}[h[H]] \, .
\eeq
We now define the classical effective average action for the point-particles  as
\beq \label{eq:gamma}
S_{k} [\bar{g}+ H,x_1,x_2] \equiv \S_{\mathrm{pp},k}[h_k[H],x_1,x_2] \, .
\eeq
Its flow is given by
\[
 \partial_k S[\bar{g}+ H]
=  \partial_k h_b[H] \, \S_{\mathrm{pp}}^{b}[h[H]] +  \partial_k  \S_{\mathrm{pp}}[h[H]] \, .
\]
Using \eqref{eq:hflow} and \eqref{eq:sflow}, this becomes
\begin{eqnarray}
 \partial_k S[\bar{g}+ H]  
&=& -   \S_{\mathrm{pp}}^{a} \, \mathcal{G}_{ac} \dot{R}^{cd} \mathcal{G}_{db} \S_{-1}^{b}
+   \S_{\mathrm{pp}}^{a} \, \mathcal{G}_{ac} \dot{R}^{cd} \mathcal{G}_{db} \S_{-1}^{b} \nonumber
+ \frac{\kappa}{2} \,  \S_{\mathrm{pp}}^{a} \, \mathcal{G}_{ac} \dot{R}^{cd} \mathcal{G}_{db}  \S_{\mathrm{pp}}^{b} \, .
\end{eqnarray}
The first two terms cancel. Using the definition of $S$ together with the first relation in \eqref{eq:hH}, we obtain from the chain rule 
\begin{align} \label{eq:flowWett2}
 \partial_k S[\bar{g}+ H]  
&= -  \frac{\kappa}{2} \, S^{(1)}[\bar{g}+ H] \cdot
\partial_k \frac{1}{S_{-1}^{(2)}[H] + R_k} 
\cdot S^{(1)}[\bar{g}+ H]  \,.
\end{align}
This expression is nearly the desired flow equation. To show that \eqref{eq:flowWett2} coincides with \eqref{eq:flowWett}, we only need to establish the equivalence between $S_{-1}^{(2)}[H]$ and $S_{\rm g}^{(2)}[\bar{g} + H]$, which indeed holds for all $k$.
%
To see this, first observe that  \eqref{eq:ballena} implies  $\lim_{k\to \infty}h_k[H]=H$. Inserting this into \eqref{eq:Gammahat} gives
\beq \notag
\lim_{k \to \infty} S_{-1}[H]
= \tfrac{1}{2} H_a \Delta^{ab} H_b + \lim_{k \to \infty}\S_{-1}[H] 
=S_{\rm g}[\bar{g}+H] \, .
\eeq  
Where we haver used the fact that $\lim_{k\to\infty} \S_{-1}[H]$ contains all cubic and higher order
gravitational vertices, while adding the quadratic term 
$\tfrac{1}{2} H_a \Delta^{ab} H_b$ reconstructs the full gravitational
action $S_g$.  Since $S_{-1}$ does not flow, it then follows that
\beq \notag
S_{-1}[H] = S_{g}[\bar{g}+H] \, 
\eeq  
independently of $k$.
Finally, replacing $\bar{g} + H$ with the full metric $g$, Eq.~\eqref{eq:flowWett2} becomes
\beq \label{eq:Wettfinal}
\boxed{\;  \partial_kS_k[g]  
= - \frac{\kappa}{2} \, S_k^{(1)}[g] \cdot 
\partial_k \frac{1}{S_{g}^{(2)}[g] + R_k} \cdot  
S_k^{(1)}[g] \; }
\eeq  
which is exactly \eqref{eq:flowWett}.

To check that this equation reproduces the correct limit for $S_k[\bar{g}]$ as $k \to 0$, recall its definition in \eqref{eq:gamma}. From this we have
\[
\lim_{k \to 0} S_k [\bar{g} + H] = \lim_{k \to 0}\S_{\mathrm{pp}, \, k}[h_k[H],x_1,x_2] .
\]
We therefore need the limits of $\S_{\mathrm{pp}}$ and the relation $h_k[H]$. Taking $k \to 0$ in \eqref{eq:raya} and evaluating it at $H=0$, we obtain $\Delta h=0$. With suitable boundary conditions, this ensures that $\lim_{k \to 0} h_k[H=0]=0$. For $\S_{\mathrm{pp}}$, from \eqref{eq:Seff_k_0} one concludes that
\[
\lim_{k \to 0} \S_{\mathrm{pp}, \, k}[0,x_1,x_2]=S_{\rm eff}[x_1,x_2]-\lim_{k \to 0} \frac{1}{\kappa}\S_{-1, \, k}[0]\, .
\]
Furthermore from \eqref{eq:Gammahat} it follows  that $\lim_{k \to 0} \S_{-1}[0]=\lim_{k \to 0} S_{-1}[0]=S_{g}[\bar{g}] = 0$, where we have used the fact thar the Einstein-Hilbert action vanishes for $R_{\mu\nu} = 0$. Hence we have that
\beq \notag
\lim_{k \to 0} S_k [\bar{g}]=S_{\rm eff}[x_1,x_2] \, .
\eeq  
In other words, solving \eqref{eq:Wettfinal} allows us to recover the effective action $S_{\rm eff}$.
Unlike the Wilsonian effective action, the initial condition for the average effective action $S_k$ is given by the point-particle action $\lim_{k\to \infty}S_k = S_{\rm pp}$, without the pure gravity part. In essence, by making the Legendre transformation, we have resummed the ``pure gravity contributions'' to the flow equation.
\\

\section{Conclusion and Outlook
}
\label{Sec:4}
\vspace{-0.2cm}

In this work, we have provided a rigorous formal foundation for the classical renormalization group (RG) approach to the general relativistic two-body problem introduced in \cite{Gutierrez:2025rqp}. By proving the exact equivalence of our heuristic flow equation \eqref{eq:flowWett} with the classical Polchinski equation \eqref{eq:floweq}, via the duality \eqref{eq:Gammahat}, we have solidified it as a powerful and non-perturbative tool for deriving the gravitational effective action.
\\

With these theoretical foundations firmly established, the path is now clear for focused applications. Our immediate future work will concentrate on developing systematic approximation schemes to solve the flow equation and extract physical predictions. We are currently developing a Post-Newtonian Derivative Expansion (PNDE) for the effective action $S_{\rm eff}[x_1,x_2]$ of the two-body system \cite{Gutierrez:2026b}. The success of this approximation will be gauged by its ability to reproduce gauge-invariant physical observables, most notably the conserved energy of the binary orbit as a function of orbital frequency \cite{Blanchet:2013PN}, and the scattering angle as a function of the impact parameter \cite{Driesse:2024feo}. A parallel line of investigation involves constructing a fully covariant ansatz for the effective action featuring a running metric and an extension of the point-particle action \cite{Porto:2016pyg}. 
%
We are optimistic that these strategies will yield efficient new methods for computing gauge invariant quantities which can rival effective one body (EOB) \cite{Buonanno:1998EOB} and more generally agree well with ``exact'' Numerical Relativity \cite{Gourgoulhon:2001ec,Grandclement:2001ed} but at a negligible computational cost.
\\

We expect that, as is the case for statistical/quantum RG equations, the effective average action formulism will turn out to be more useful than the Wilsonian one in practice. Indeed, there are analytical arguments that suggest that the derivative expansion of the effective average action has a finite radius of convergence \cite{Balog:2019rrg}, which do not apply to the Polchinski equation. Lastly, let us comment that gravity is only one possible application of our classical RG equation; there may exist many other fields of research where our equation can be applied.

\section*{Acknowledgments}
\vspace{-0.2cm}


The authors acknowledge financial support from the CSIS grant I+D-2022 2252 0220 1001 74UD. 
A.C. also acknowledges financial support from ANII SNI 2023 1 1013433. 
F.G. acknowledges support from the Comisión Académica de Posgrados CAP, Universidad de la República (Uruguay), through a Master's fellowship, 
and gratefully acknowledges the hospitality and support of the Ca’ Foscari University of Venice during a research internship.
\\

\appendix

\section{Post-Minkowskian Expansion from the Classical Polchinski Equation}

In this section we show that also in the Polchinski framework for classical gravity it is possible to reproduce perturbation theory -- up to at least order $\kappa^3$ (3PM). The equivalent derivation in the average action framework was given already in \cite{Gutierrez:2025rqp}. Since the two approaches differ in subtle yet instructive ways, we provide here the Polchinski variant as well.
 In this section $S_{\rm pp}$ and its derivatives, $S_{\rm pp}^a$ etc., are evaluated at $g= \bar{g}$.
\\

For convenience, we suppress the $k$ dependence, writing $\mathcal{G} \equiv \mathcal{G}_k$ and 
$\S \equiv \S_k$, while $S_g$ and $S_{\mathrm{pp}}$ remain $k$-independent. 
We begin by expanding the action $\S$ as
\[
\S = \frac{1}{\kappa} \S_{-1} + \S_{0} + \kappa \S_1 + \kappa^2 \S_2+ \kappa^3 \S_3 + O(\kappa^4)
\]
and substitute this into the flow equation \eqref{eq:floweq}, obtaining
\begin{align*}
\frac{1}{\kappa}\,k\partial_k \S_{-1}
&+ k\partial_k \S_{0}
+ \kappa\,k\partial_k \S_{1}
+ \kappa^{2} k\partial_k \S_{2}
\nonumber\\[2pt]
&=- \frac{\kappa}{2} \Bigl( \tfrac{1}{\kappa}\S_{-1}^{a} + \S_{0}^{a} + \kappa \S_{1}^{a} + \kappa^{2}\S_{2}^{a} \Bigr) 
\,\dot{\mathcal{G}}_{ab}\, 
\Bigl( \tfrac{1}{\kappa}\S_{-1}^{b} + \S_{0}^{b} + \kappa \S_{1}^{b} + \kappa^{2}\S_{2}^{b} \Bigr) 
+ \mathcal{O}(\kappa^{3}) \,.
\end{align*}
This expression can be reorganised as
\begin{align}
\frac{1}{\kappa} \, k\partial_k \S_{-1} 
&+ k\partial_k \S_{0} 
+ \kappa\, k\partial_k \S_{1} 
+ \kappa^{2} k\partial_k \S_{2} 
\nonumber\\[4pt]
&=- \tfrac{1}{2\kappa}\,\S_{-1}^{a}\,\dot{\mathcal{G}}_{ab}\,\S_{-1}^{b} 
- \S_{0}^{a}\,\dot{\mathcal{G}}_{ab}\,\S_{-1}^{b} 
- \tfrac{\kappa}{2}\,\S_{0}^{a}\,\dot{\mathcal{G}}_{ab}\,\S_{0}^{b} 
\nonumber\\[4pt]
&\quad -\;\kappa\,\S_{1}^{a}\,\dot{\mathcal{G}}_{ab}\,\S_{-1}^{b} 
-\kappa^{2}\,\S_{0}^{a}\,\dot{\mathcal{G}}_{ab}\,\S_{1}^{b} 
-\kappa^{2}\,\S_{-1}^{a}\,\dot{\mathcal{G}}_{ab}\,\S_{2}^{b} 
+ \mathcal{O}(\kappa^{3}) \,.
\label{eq:pertflow}
\end{align}

\subsubsection*{Calculation of $\S_{-1}$}

Projecting the $\kappa^{-1}$ sector of \eqref{eq:pertflow} gives
\beq\label{eq:-1flow}
k\partial_k \S_{-1} = -\frac{1}{2} \S_{-1}^{a}\,\dot{\mathcal{G}}_{ab}\,\S_{-1}^{b} \,.
\eeq
We denote by $F[0]$ the evaluation of $F$ at $h=0$, so that $\S_{-1}[0]=\S_{-1}|_{h=0}$. Then by taking a functional derivative of \eqref{eq:-1flow} one gets to
\[
k\partial_k \S_{-1}^{a}[0] = -\S_{-1}^{ca}[0]\,\dot{\mathcal{G}}_{cb}\,\S_{-1}^{b}[0] \,.
\]
From the initial condition \eqref{eq:IC} and \eqref{eq:BEoM} we have $\S_{-1}^{a}[0]|_{k \to \infty} = S_{g}^{a}[\bar{g}]=0$. 
Since $\S_{-1}^{a}[0]=0$ is a fixed point of the flow, it follows that
\begin{equation}\label{eq:jirafaf}
\S_{-1}^{a}[0] = 0
\end{equation}
for all $k$. In particular, \eqref{eq:-1flow} at $h=0$ reduces to $k\partial_k\S_{-1}[0]=0$, which together with the initial condition \eqref{eq:IC} implies
\[
\;\S_{-1}[0] = S_g[\bar{g}]\;
\]
for all $k$. The effective action can be computed remembering that $S_{\rm eff}[x_1,x_2] = \lim_{k\to 0} \S_k[0,x_1,x_2]$, so one gets to 
\[
\boxed{\;S_{{\rm eff},-1} = S_g[\bar{g}]\;}
\]
Now we proceed by finding the higher point functions $\S_{-1}^{ab}[0]$, $\S_{-1}^{abc}[0]$ and $\S_{-1}^{abcd}[0]$ which are going to be needed to find the following orders of $S_{{\rm eff}}$.  Differentiating \eqref{eq:-1flow} once more yields
\begin{align*}
k\partial_k \S_{-1}^{ab}[0] 
&=- \S_{-1}^{ac}[0]\dot{\mathcal{G}}_{cd}\,\S_{-1}^{db}[0]
- \S_{-1}^{abc}[0]\dot{\mathcal{G}}_{cd}\,\cancelto{0}{\S_{-1}^{d}[0]} \\
&=-\S_{-1}^{ac}[0]\dot{\mathcal{G}}_{cd}\,\S_{-1}^{db}[0] 
\end{align*}
where we used \eqref{eq:jirafaf}. Again, $\S_{-1}^{ab}[0]=0$ is a fixed point, and since $\lim_{k\to\infty}\S_{-1}^{ab}[0]=0$, we conclude
\beq\label{eq:rinoceronte}
\S_{-1}^{ab}[0] = 0 \,.
\eeq
Proceeding analogously, one finds that
\beq \label{eq:cierva}
\S_{-1}^{abc}[0] = S_{g}^{abc}[\bar{g}] \,,
\eeq
while for the four-point term one obtains
\[
k\partial_k \S_{-1}^{abcd}[0] = - S_{g}^{abe}[\bar{g}] \dot{\mathcal{G}}_{ef} S_{g}^{fcd}[\bar{g}]
- S_{g}^{ace}[\bar{g}] \dot{\mathcal{G}}_{ef} S_{g}^{fbd}[\bar{g}]
- S_{g}^{ade}[\bar{g}] \dot{\mathcal{G}}_{ef} S_{g}^{fbc}[\bar{g}] \,.
\]
Integrating this equation and using the initial condition $\lim_{k \to \infty} \S_{-1}^{abcd}[0]= S_{g}^{abcd}[\bar{g}]$ we obtain
\beq \label{eq:carpincho}
\begin{aligned}
\S_{-1}^{abcd}[0]
&= S_{g}^{abcd}[\bar{g}] - S_{g}^{abe}[\bar{g}]\, \mathcal{G}_{ef}\, S_{g}^{fcd}[\bar{g}]  \\
&\quad- S_{g}^{ace}[\bar{g}]\, \mathcal{G}_{ef}\, S_{g}^{fbd}[\bar{g}]- S_{g}^{ade}[\bar{g}]\, \mathcal{G}_{ef}\, S_{g}^{fbc}[\bar{g}]  \, ,
\end{aligned}
\eeq
with $\mathcal{G}$ given by \eqref{eq:cebra}.

\subsubsection*{Calculation of $\S_{0}$}

Let us now consider the $\kappa^0$ sector of \eqref{eq:pertflow}, which yields
\beq \label{eq:s0}
k\partial_k \S_0 = - \S_{0}^{a}\,\dot{\mathcal{G}}_{ab}\,\S_{-1}^{b} \,.
\eeq
Using \eqref{eq:jirafaf}, this immediately implies
\[
k\partial_k \S_0[0]=0 \,,
\]
and from the initial condition \eqref{eq:IC} it follows that
\beq \nonumber
\boxed{\;S_{{\rm eff},0}= S_{\mathrm{pp}}\;}
\eeq
for all $k$.
Differentiating \eqref{eq:s0} gives
\[
k\partial_k \S_{0}^{a}[0] 
= -\S_{0}^{ab}[0]\,\dot{\mathcal{G}}_{bc}\,\S_{-1}^{c}[0]
- \S_{0}^{b}[0]\,\dot{\mathcal{G}}_{bc}\,\S_{-1}^{ac}[0] \,.
\]
Using \eqref{eq:jirafaf} and \eqref{eq:rinoceronte}, this reduces to 
\[
k\partial_k \S_{0}^{a}[0]=0 \,.
\]
Thus, from the initial condition we obtain
\beq \label{eq:tigre}
\S_{0}^{a}[0] = S_{\mathrm{pp}}^{a} \,.
\eeq
Proceeding further, one finds
\begin{align*}
k\partial_k \S_{0}^{ab}[0] 
&= - \S_{0}^{abc}[0] \,\dot{\mathcal{G}}_{cd}\, \cancelto{0}{\S_{-1}^{d}[0]} - \S_{0}^{ac}[0] \,\dot{\mathcal{G}}_{cd}\, \cancelto{0}{\S_{-1}^{db}[0]}\\
&\quad - \S_{0}^{bc}[0] \,\dot{\mathcal{G}}_{cd}\, \cancelto{0}{\S_{-1}^{ad}[0]}- \S_{0}^{d}[0] \,\dot{\mathcal{G}}_{dc}\, \S_{-1}^{abc}[0] \, .
\end{align*}
Then, using \eqref{eq:cierva}, this reduces to
\beq \nonumber
k\partial_k \S_{0}^{ab}[0] = - S_{\mathrm{pp}}^{c} \,\dot{\mathcal{G}}_{cd}\, S_{g}^{dab} \,.
\eeq
Integrating, with the initial condition $\lim_{k\to\infty}\S_{0}^{ab}[0]=S_{\mathrm{pp}}^{ab}$, gives
\beq \label{eq:zorro}
\S_{0}^{ab}[0] = S_{\mathrm{pp}}^{ab} - S_{\mathrm{pp}}^{c}\,\mathcal{G}_{cd}\,S_{g}^{dab} \,.
\eeq
If then one continues with $\S_{0}^{abc}[0]$ one has
\begin{align*}
k \partial_k \S_{0}^{abc}[0] 
&= - \S_{0}^{ad}[0] \,\dot{\mathcal{G}}_{de}\, \S_{-1}^{ebc}[0]- \S_{0}^{bd}[0] \,\dot{\mathcal{G}}_{de}\, \S_{-1}^{eac}[0] \\
&\quad - \S_{0}^{cd}[0] \,\dot{\mathcal{G}}_{de}\, \S_{-1}^{eab}[0]- \S_{0}^{d}[0]  \,\dot{\mathcal{G}}_{de}\, \S_{-1}^{eabc}[0]  \, .
\end{align*}
By using  \eqref{eq:cierva}, \eqref{eq:carpincho},  \eqref{eq:tigre} and \eqref{eq:zorro} we arrive at 
\begin{align} 
\S_{0}^{abc}[0] &= S_{\mathrm{pp}}^{abc} -S_{\mathrm{pp}}^{d}\mathcal{G}_{de}S_{g}^{eabc}[\bar{g}] \nonumber \\ 
&-\left(S_{\mathrm{pp}}^{ad} \mathcal{G}_{de}S_{g}^{ebc}[\bar{g}]+S_{\mathrm{pp}}^{bd} \mathcal{G}_{de}S_{g}^{eac}[\bar{g}]+S_{\mathrm{pp}}^{cd} \mathcal{G}_{de}S_{g}^{eab}[\bar{g}] \right) \nonumber \\
&+\left(S_{g}^{bce}[\bar{g}] \mathcal{G}_{de}S_{g}^{efa}[\bar{g}] \mathcal{G}_{fh}S_{\mathrm{pp}}^{h}+S_{g}^{ace}[\bar{g}] \mathcal{G}_{de}S_{g}^{efb}[\bar{g}] \mathcal{G}_{fh}S_{\mathrm{pp}}^{h} \right.\nonumber\\
&\left.+S_{g}^{abe}[\bar{g}] \mathcal{G}_{de}S_{g}^{efc}[\bar{g}] \mathcal{G}_{fh}S_{\mathrm{pp}}^{h} \right) \label{eq:trid}\,.
\end{align}

\subsubsection*{Calculation of $\S_{1}$}

The $\kappa^1$ contribution to \eqref{eq:pertflow} reads
\beq \label{eq:1PM}
k\partial_k \S_1 = - \frac{1}{2}\,\S_{0}^{a}\,\dot{\mathcal{G}}_{ab}\,\S_{0}^{b} - \S_{1}^{a}\,\dot{\mathcal{G}}_{ab}\,\S_{-1}^{b} \,.
\eeq
Evaluating at $h=0$ and using \eqref{eq:jirafaf} and \eqref{eq:tigre}, we find
\[
k\partial_k \S_1[0] = -k\partial_k \left( \tfrac{1}{2}\, S_{\mathrm{pp}}^{a}\,\mathcal{G}_{ab}\,S_{\mathrm{pp}}^{b} \right) \,.
\]
Since $\S_1[0]\to 0$ as $k\to\infty$, integration yields
\[
\S_{1}[0] = - \tfrac{1}{2}\, S_{\mathrm{pp}}^{a}\,G_{ab}\,S_{\mathrm{pp}}^{b} \,.
\]
Taking the limit $k\to 0$, where $G_{ab}\to (\Delta^{-1})_{ab}$, we obtain
\[
\boxed{\,S_{{\rm eff},1} = - \tfrac{1}{2}\, S_{\mathrm{pp}}^{a}\,(\Delta^{-1})_{ab}\,S_{\mathrm{pp}}^{b} \,}
\]
which is the expected result.

To determine $S_{{\rm eff},2}$, the flow of $\S_{1}^{a}[0]$ will be needed. Differentiating \eqref{eq:1PM} gives
\[
k\partial_k \S_{1}^{a}[0] = -\S_{0}^{ab}[0]\,\dot{\mathcal{G}}_{bc}\,S_{\mathrm{pp}}^{c} \,,
\]
and using \eqref{eq:zorro} this becomes
\begin{align*}
k\partial_k \S_{1}^{a}[0] 
&= -\left(S_{\mathrm{pp}}^{ab} - S_{\mathrm{pp}}^{c}\,\mathcal{G}_{cd}\,S_{g}^{dab}\right) \dot{\mathcal{G}}_{bc}\,S_{\mathrm{pp}}^{c} \\
&= -S_{\mathrm{pp}}^{ab}\dot{\mathcal{G}}_{bc}\,S_{\mathrm{pp}}^{c} + S_{\mathrm{pp}}^{e}\,\mathcal{G}_{ed}\,S_{g}^{dab}\,\dot{\mathcal{G}}_{bc}\,S_{\mathrm{pp}}^{c} \\
&= k\partial_k \left( -S_{\mathrm{pp}}^{ab}\,\mathcal{G}_{bc}\,S_{\mathrm{pp}}^{c} + \tfrac{1}{2} S_{\mathrm{pp}}^{e}\,\mathcal{G}_{ed}\,S_{g}^{dab}\,\mathcal{G}_{bc}\,S_{\mathrm{pp}}^{c} \right) \, .
\end{align*}
With the initial condition $\S_{1}^{a}[0]\to 0$ as $k\to\infty$, we integrate to obtain
\beq \label{eq:S1a}
\S_{1}^{a}[0] = - S_{\mathrm{pp}}^{ab}\,\mathcal{G}_{bc}\,S_{\mathrm{pp}}^{c} 
+ \tfrac{1}{2}\, S_{\mathrm{pp}}^{e}\,\mathcal{G}_{ed}\,S_{g}^{dab}[\bar{g}]\,\mathcal{G}_{bc}\,S_{\mathrm{pp}}^{c} \,.
\eeq
To evaluate $S_{\mathrm{eff},3}$ we also require the function 
$\S_{-1}^{ab}[0]$. Differentiating \eqref{eq:1PM} once again yields
\[
k\partial_k\S_{1}^{ab}[0] = - \S_{0}^{abc}[0] \,  \dot{\mathcal{G}}_{cd} \, \S_{0}^{d}[0]- \S_{0}^{ac}[0]\,\dot{\mathcal{G}}_{cd} \, \S_{0}^{bd}[0]- \S_{1}^{c}[0] \, \dot{\mathcal{G}}_{cd} \, \S_{-1}^{abd}[0]\,.
\]
Using \eqref{eq:carpincho}, \eqref{eq:tigre}, \eqref{eq:zorro}, \eqref{eq:trid}  and \eqref{eq:S1a}, we obtain
\begin{align} \label{eq:S1ab}
\S_{1}^{ab}[0] &= \tfrac{1}{2} S_{g}^{abcd}[\bar{g}] \, \mathcal{G}_{ce} \mathcal{G}_{df} \, S_{\mathrm{pp}}^{e} \, S_{\mathrm{pp}}^{f} 
- S_{\mathrm{pp}}^{ac} \mathcal{G}_{cd} S_{\mathrm{pp}}^{db} \nonumber \\ 
&\quad - S_{g}^{acd}[\bar{g}] \, \mathcal{G}_{de} S_{g}^{bef}[\bar{g}] \, \mathcal{G}_{ci} S_{\mathrm{pp}}^{i} \, \mathcal{G}_{fk} S_{\mathrm{pp}}^{k} 
+ S_{\mathrm{pp}}^{abc} \, \mathcal{G}_{cd} S_{\mathrm{pp}}^{d} \nonumber \\
&\quad + S_{g}^{acd}[\bar{g}] \mathcal{G}_{ce} S_{\mathrm{pp}}^{be} \mathcal{G}_{df} S_{\mathrm{pp}}^{f} 
+ S_{g}^{abc}[\bar{g}] \mathcal{G}_{cd} S_{\mathrm{pp}}^{de} \mathcal{G}_{ef} S_{\mathrm{pp}}^{f}\nonumber \\
&\quad- \tfrac{1}{2} S_{g}^{abc}[\bar{g}] \mathcal{G}_{cd} S_{g}^{def}[\bar{g}] \mathcal{G}_{eh} S_{\mathrm{pp}}^{h} \mathcal{G}_{fi} S_{\mathrm{pp}}^{i}  \, .
\end{align}

\subsubsection*{Calculation of $\S_{2}$}

Proceeding to the $\kappa^2$ term, \eqref{eq:pertflow} yields
\begin{equation} \label{eq:flowS2}
k\partial_k \S_2 = - \S_{0}^{a}\,\dot{\mathcal{G}}_{ab}\,\S_{1}^{b} - \S_{-1}^{a}\,\dot{\mathcal{G}}_{ab}\,\S_{2}^{b} \,.
\end{equation}
At $h=0$, this reduces to
\[
k\partial_k \S_2[0] = - S_{\mathrm{pp}}^{a}\,\dot{\mathcal{G}}_{ab}\,\S_{1}^{b}[0] \,.
\]
Substituting the expression for $\S_{1}^{b}[0]$ of \eqref{eq:S1a}, one obtains
\begin{align*}
k\partial_k \S_2[0] &= S_{\mathrm{pp}}^{f}\,\dot{\mathcal{G}}_{fa}\,S_{\mathrm{pp}}^{b}\,\mathcal{G}_{bc}\,S_{\mathrm{pp}}^{ca} \\
&\quad - \tfrac{1}{2} S_{\mathrm{pp}}^{f}\,\dot{\mathcal{G}}_{fa}\,S_{\mathrm{pp}}^{b}\,\mathcal{G}_{bc}\,S_{\mathrm{pp}}^{e}\,\mathcal{G}_{ed}\,S_{g}^{dac}[\bar{g}] \\
&= k\partial_k \Bigl( \tfrac{1}{2}\, S_{\mathrm{pp}}^{f}\,\mathcal{G}_{fa}\,S_{\mathrm{pp}}^{b}\,\mathcal{G}_{bc}\,S_{\mathrm{pp}}^{ca} \\
&\qquad - \tfrac{1}{6}\, S_{\mathrm{pp}}^{f}\,\mathcal{G}_{fa}\,S_{\mathrm{pp}}^{b}\,\mathcal{G}_{bc}\,S_{\mathrm{pp}}^{e}\,\mathcal{G}_{ed}\,S_{g}^{dac}[\bar{g}] \Bigr).
\end{align*}
Since $\S_2[0]\to 0$ as $k\to\infty$, integration gives
\[
\fbox{%
$\begin{aligned}
\;S_{{\rm eff},2} &= \tfrac{1}{2}\, S_{\mathrm{pp}}^{f}\,(\Delta^{-1})_{fa}\,S_{\mathrm{pp}}^{b}\,(\Delta^{-1})_{bc}\,S_{\mathrm{pp}}^{ca} \\
&\quad - \tfrac{1}{6}\, S_{\mathrm{pp}}^{f}\,(\Delta^{-1})_{fa}\,S_{\mathrm{pp}}^{b}\,(\Delta^{-1})_{bc}\,S_{\mathrm{pp}}^{e}\,(\Delta^{-1})_{ed}\,S_{g}^{dac}[\bar{g}] \;
\end{aligned}$}
\]
To find $S_{{\rm eff},3}$, we will also need $\S_{2}^{a}[0]$. Differentiating \eqref{eq:flowS2} gives
\[
k\partial_k \S_{2}^{a}[0] = - \S_{0}^{ab}[0]\,\dot{\mathcal{G}}_{bc}\,\S_{1}^{c}[0] - \S_{0}^{b}[0]\,\dot{\mathcal{G}}_{bc}\,\S_{1}^{ac}[0] \,.
\]
Using \eqref{eq:tigre}, \eqref{eq:zorro},\eqref{eq:S1a} and \eqref{eq:S1ab} we get to
\begin{align} \label{eq:S2a}
\S_{2}^{a}[0]&=S_{\mathrm{pp}}^{ab}\mathcal{G}_{bc}S_{\mathrm{pp}}^{cd}\mathcal{G}_{de}S_{\mathrm{pp}}^{e}-\tfrac{1}{2}S_{\mathrm{pp}}^{ab}\mathcal{G}_{bc}S_{g}^{cde}[\bar{g}]\mathcal{G}_{df}S_{\mathrm{pp}}^{f}\mathcal{G}_{eh}S_{\mathrm{pp}}^{h} 
\\
&-S_{g}^{abc}[\bar{g}]\mathcal{G}_{bd}S_{\mathrm{pp}}^{d}\mathcal{G}_{ce}S_{\mathrm{pp}}^{ef}\mathcal{G}_{fh}S_{\mathrm{pp}}^{h}+\tfrac{1}{2}S_{\mathrm{pp}}^{abc}\mathcal{G}_{bd}S_{\mathrm{pp}}^{d}\mathcal{G}_{ce}S_{\mathrm{pp}}^{e}  \nonumber
\\
&+\tfrac{1}{2}S_{g}^{abc}[\bar{g}]\mathcal{G}_{bd}S_{\mathrm{pp}}^{d}\mathcal{G}_{ce}S_{g}^{efh}[\bar{g}]\mathcal{G}_{fi}S_{\mathrm{pp}}^{i}\mathcal{G}_{hj}S_{\mathrm{pp}}^{j} \nonumber
\\
&-\tfrac{1}{6}S_{g}^{abcd}[\bar{g}]\mathcal{G}_{be}S_{\mathrm{pp}}^{e}\mathcal{G}_{cf}S_{\mathrm{pp}}^{f}\mathcal{G}_{dh}S_{\mathrm{pp}}^{h} \nonumber\, .
\end{align}

\subsubsection*{Calculating $\S_{3}$}

Finally, the $\kappa^3$ part of \eqref{eq:pertflow} gives
\beq \nonumber
  k\partial_k \S_3 =  -\S_{3}^{a} \dot{\mathcal{G}}_{ab} \S_{-1}^{b} - \S_{2}^{a} \dot{\mathcal{G}}_{ab} \S_{0}^{b}-\frac{1}{2}\S_{1}^{a} \dot{\mathcal{G}}_{ab} \S_{1}^{b}
\eeq
at $h=0$ we get 
\begin{align*}
  k\partial_k \S_3[0] &=  -\S_{3}^{a}[0] \dot{\mathcal{G}}_{ab} \cancelto{0}{\S_{-1}^{b}[0]} - \S_{2}^{a}[0] \dot{\mathcal{G}}_{ab} \S_{0}^{b}[0]-\frac{1}{2}\S_{1}^{a}[0] \dot{\mathcal{G}}_{ab} \S_{1}^{b}[0]\\
  &=- \S_{2}^{a}[0] \dot{\mathcal{G}}_{ab} S_{\mathrm{pp}}^{b}-\frac{1}{2}\S_{1}^{a}[0] \dot{\mathcal{G}}_{ab} \S_{1}^{b}[0]\,.
  \end{align*}
By introducing \eqref{eq:S1a} and \eqref{eq:S2a} and integrating up to $k=0$ we get the final result
{
\setlength{\fboxsep}{8pt}
\setlength{\fboxrule}{0.6pt}
\setlength{\jot}{8pt} 
\begin{empheq}[box=\fbox]{align*}
  S_{{\rm eff},3} &= \tfrac{1}{24} S_{g}^{abcd}[\bar{g}] (\Delta^{-1})_{ae} S_{\mathrm{pp}}^{e} (\Delta^{-1})_{bf} S_{\mathrm{pp}}^{f} 
            (\Delta^{-1})_{ch} S_{\mathrm{pp}}^{h} (\Delta^{-1})_{di} S_{\mathrm{pp}}^{i} \nonumber \\
  &\quad - \tfrac{1}{8} S_{g}^{abc}[\bar{g}] (\Delta^{-1})_{ad} S_{g}^{def}[\bar{g}] (\Delta^{-1})_{bh} S_{\mathrm{pp}}^{h} 
            (\Delta^{-1})_{ci} S_{\mathrm{pp}}^{i} (\Delta^{-1})_{ej} S_{\mathrm{pp}}^{j} (\Delta^{-1})_{fk} S_{\mathrm{pp}}^{k} \nonumber \\
  &\quad + \tfrac{1}{2} S_{\mathrm{pp}}^{a} (\Delta^{-1})_{ab} S_{\mathrm{pp}}^{bc} (\Delta^{-1})_{cd} S_{g}^{def}[\bar{g}] 
            (\Delta^{-1})_{eh} S_{\mathrm{pp}}^{h} (\Delta^{-1})_{fi} S_{\mathrm{pp}}^{i} \nonumber \\
  &\quad - \tfrac{1}{2} S_{\mathrm{pp}}^{a} (\Delta^{-1})_{ab} S_{\mathrm{pp}}^{bc} (\Delta^{-1})_{cd} S_{\mathrm{pp}}^{de} 
            (\Delta^{-1})_{ef} S_{\mathrm{pp}}^{f} \nonumber \\
  &\quad - \tfrac{1}{6} S_{\mathrm{pp}}^{abc} (\Delta^{-1})_{ad} S_{\mathrm{pp}}^{d} (\Delta^{-1})_{be} S_{\mathrm{pp}}^{e} 
            (\Delta^{-1})_{cf} S_{\mathrm{pp}}^{f} 
\end{empheq}
}
This concludes the determination of the effective action to order 3PM.

\end{document}